# Modeling of micro- and nano-scale domain recording by high-voltage atomic force microscopy in ferroelectrics-semiconductors.

Anna N. Morozovska[1], Eugene A. Eliseev[2].

[1]V.Lashkaryov Institute of Semiconductor Physics, National Academy of Science of Ukraine,

41, pr. Nauki, 03028 Kiev, Ukraine,

e-mail: morozo@i.com.ua

[2]Institute for Problems of Materials Science, National Academy of Science of Ukraine,

3, Krjijanovskogo, 03142 Kiev, Ukraine

e-mail: eliseev@i.com.ua

**ABSTRACT**

The equilibrium sizes of micro- and nano-domains caused by electric field of atomic force microscope tip in ferroelectric semiconductor crystals have been calculated. The domain was considered as a prolate semi-ellipsoid with rather thin domain walls. For the first time we modified the Landauer model allowing for semiconductor properties of the sample and the surface energy of the domain butt. The free carriers inside the crystal lead to the formation of the screening layer around the domain, which partially shields its interior from the depolarization field. We expressed the radius and length of the domain though the crystal material parameters (screening radius, spontaneous polarization value, dielectric permittivity tensor) and atomic force microscope tip characteristics (charge, radius of curvature). The obtained dependence of domain radius via applied voltage is in a good quantitative agreement with the ones of submicron ferroelectric domains recorded by high-voltage atomic force and scanning probe microscopy in $LiNbO_3$ and $LiTaO_3$ crystals.

## 1. INTRODUCTION

Submicron spatial regions with reversed spontaneous polarization called micro- and nano- ferroelectric domains have been observed in many ferroelectrics [1-5]. It is clear from general point of view that domains formation can be caused by strong local electric fields with definite polarity. In particular they could originate near microscopic charged defects, dislocations, micro-aggregates, surface defects and so on.

Recently one and two dimensional arrays of spike-like nano-domains have been fabricated in $LiNbO_3$ [1], $LiTaO_3$ [2], $Pb(Zr,Ti)O_3$ [3], $BaTiO_3$ [4], $RbTiOPO_4$ and $RbTiOAsO_4$ [5] ferroelectric crystals with the help of electric fields caused by atomic force microscope (AFM) tip. Obtained nano-domain arrays could be successfully used in modern large-capacity memory devices and light converters based on second harmonic generation. So the possibilities of information recording in the ferroelectric media have been open, if only the optimization problem of high-speed writing nano-domains with high density, stability and fully controllable reversibility would be solved. First of all it is necessary to record the stable domain "dots" with minimum width in the appropriate ferroelectric medium. To realize this idea, one has to determine the dependences of domain radius and length on voltage applied to the AFM tip and ferroelectric medium characteristics either empirically or theoretically. To our mind simple analytical formulas for the correct description of the numerous experimental results seem rather urgent, but present phenomenological models cannot be directly applied to the nano-domain tailoring using charged AFM tip owing to the following reasons.

- The phenomenological description of the nucleation processes in the perfect dielectric during ferroelectric polarization switching proposed by Landauer [6] should be applied to the domain formation with great care, because in this model the depolarization field is partially screened by the free charges on the superconducting electrodes. When modeling the microdomains formation such upper electrode will completely screen the interior of ferroelectric from the AFM tip electric field, thus no external source would induce the polarization reversal (see Fig. 1a). Only homogeneous external field can be applied to this system.

- Theoretical modeling of ferroelectric equilibrium domains recorded by AFM tip proposed in the paper [7] considers tip electric field inside the perfect dielectric-ferroelectric with free surface, i.e. without any screening layer or upper electrode, but the semi-ellipsoidal domain depolarization field was calculated in Landauer model as if the perfect external screening expected (see Fig. 1a). As a result they obtained significantly over-estimated values of domain radius at high voltages [1].

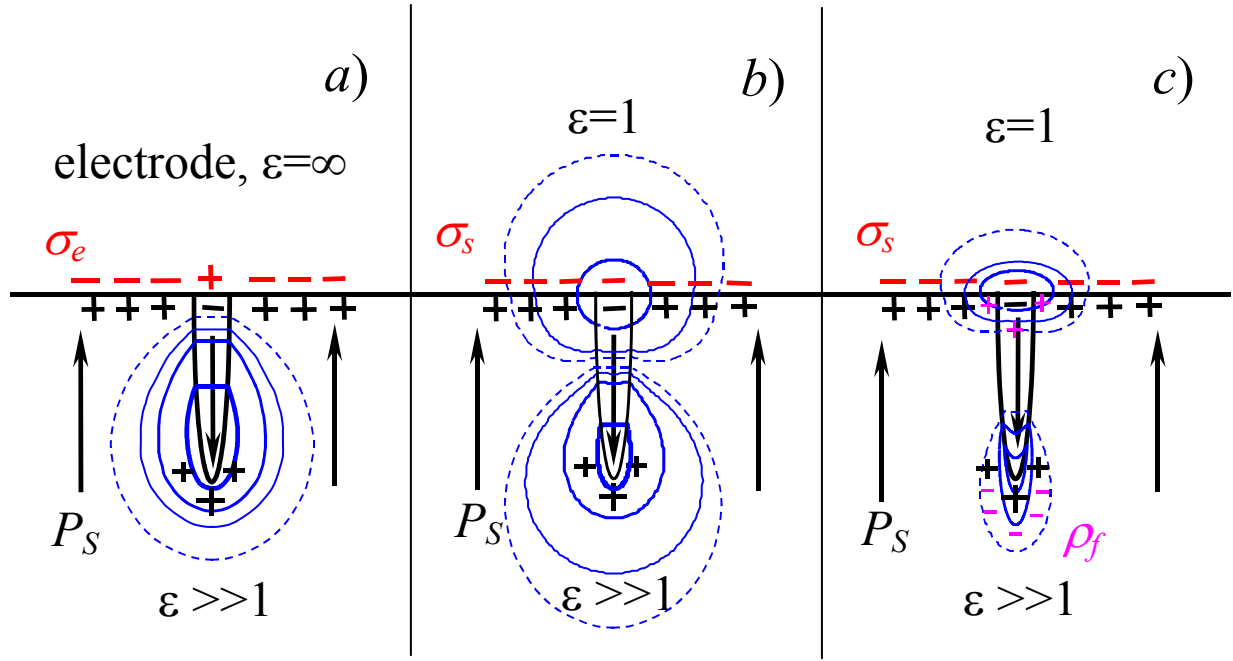

**Figure 1**. Isopotential lines of depolarization field caused by semi-ellipsoid domain for different models: Landauer model [6] with movable charges $\sigma_e$ on the ideal electrode (a), model with surface charges $\sigma_S$ captured by traps (b), model with surface charges $\sigma_S$ and bulk free charges $\rho_f$ (c).

In the present paper we try to overcome the aforementioned discrepancy, mainly taking into consideration the surface charge layers [8] and semiconductor propertied revealed by the most of ferroelectrics [9, 10]. In ferroelectric-semiconductor the Schottky barriers, band bending, field effects as well as Debye screening cause surface charge layer appearance that effectively shields the interior of the sample from the strong homogeneous depolarization field $E_d = -4\pi P_S$ [10]. Thus we assume, that carriers captured by the surface trap levels completely compensate homogeneous depolarization field existed in the absence of domain. The screening layer maintains its negative charge during the domain formation owing to the traps sluggishness and strong attraction of positively charged AFM tip. Therefore it transmits the charged tip electric field, although it does not compensate the depolarization field created by domain butt (see Fig. 1b). The free carriers inside the crystal lead to the formation of the smeared screening layer around the domain, which partially shields its interior from both the depolarization field and the external electric field caused by the positively charged AFM tip (see Fig. 1c).

The potential distribution inside the ferroelectric crystal was analyzed using a simplified rigid model [6, 7, 11] that ignores electroelastic coupling in the ferroelectric. This is possible if AFM tip does not touch the surface. Additional surface energy of the domain butt and "screening" of the depolarization and interaction energies lead to the essential decrease of the equilibrium domain sizes. As a result we obtained real values of domain radius recorded in LiNbO$_3$ and LiTaO$_3$ crystals even at high external voltages, in contrast to the over-estimated ones calculated within Landauer model evolved for the perfect polar dielectrics covered with superconducting electrodes.

## 2. PHENOMENOLOGICAL DESCRIPTION

For description of the charged tip electric field we use the point-charge model [7], in which the tip is represented by the effective point charge $q$ located at the distance $z_0$ from the sample surface. The voltage $U$ is applied between the tip and ground electrode. Then one can find the charge $q$ and charge-surface distance $z_0$ effective values from the conditions that at distance $\Delta R$ from the surface isopotential line has the tip radius of curvature $R_0$ (see [11] and Fig.2).

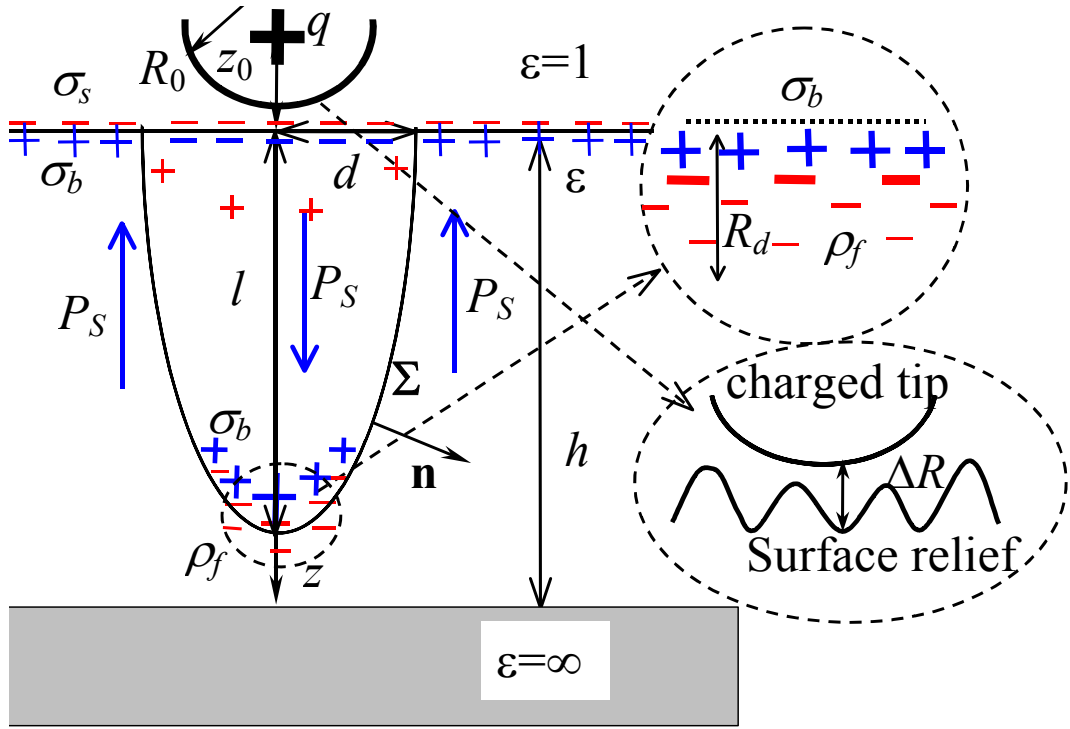

**Figure 2**. Domain formation induced by positively charged tip with effective charge $q$ and distance $z_0$ from the surface, $R_0$ is tip radius of curvature, $\Delta R$ is the distance between the tip apex and the sample surface, $d$ is semi-ellipsoid radius, $l$ is semi-ellipsoid major axis, $R_d$ is the thickness of the screening space-charge layer, $P_S$ is spontaneous polarization, $\sigma_S$ is surface charges captured on the trap levels, $\sigma_b$ is bound charges related to $P_S$ discontinuity, $\rho_f$ is free charge density. We choose constant spontaneous polarization $+P_S$ inside and $-P_S$ outside the domain. The system as a whole is electro neutral.

Hereinafter we use the model of the smeared screening layer around the semi-ellipsoidal domain region inside the rigid ferroelectric with dielectric permittivity $\varepsilon$, displacement $\mathbf{D} = \varepsilon \cdot \mathbf{E} + 4\pi \mathbf{P}_S$ and electric field $\mathbf{E} = -\nabla\varphi(\mathbf{r})$. Surface charges $\sigma_S = -P_S$ are captured on the trap levels before the domain formation. They are almost immovable during the polarization reversal. Inside an extrinsic semiconductor $|Ne\varphi(\mathbf{r})/k_B T| \ll 1$ and thus the screening of electric field $\mathbf{E} = -\nabla\varphi(\mathbf{r})$ is realized by free charges with bulk density $\rho_f(\mathbf{r}) \approx -\varepsilon\varphi(\mathbf{r})/4\pi R_d^2$ and Debye screening radius $R_d$ (see inset to the Fig.2 and Appendix A).

The potential spatial distribution caused by the tip with effective charge $q$ localized in air in the point $r_0 = (0,0,-z_0)$, and polarization reversal inside semi-ellipsoidal domain should be determined from the Maxwell equations. Maxwell equation $\Delta\varphi(\mathbf{r}) = -\dfrac{4\pi}{\varepsilon}\rho_f(\mathbf{r})$ with interfacial conditions $D_{n\,\text{int}} = D_{n\,\text{ext}}$ on the domain surface $\Sigma$, $D_{n\,\text{ext}} - D_{n\,\text{int}} = 4\pi\sigma_S$ on the free surface $z = 0$ and potential disappearance at the bottom electrode $z = h$, can be written as following boundary problem:

$$\Delta\varphi_0(\mathbf{r}) = -4\pi q\,\delta(x,y,z+z_0), \qquad z \leq 0,$$

$$\varphi_0(z=0) = \varphi(z=0), \quad \left(\frac{\partial\varphi_0}{\partial z} - \varepsilon\frac{\partial\varphi}{\partial z}\right)\bigg|_{z=0} = \begin{cases} -8\pi P_S, & \sqrt{x^2+y^2} < d \\ 0, & \sqrt{x^2+y^2} > d \end{cases}$$

$$\Delta\varphi(\mathbf{r}) - \frac{\varphi(\mathbf{r})}{R_d^2} = 0, \qquad 0 \leq z \leq h,$$

$$\varepsilon\left(\frac{\partial\varphi_{\text{int}}}{\partial n} - \frac{\partial\varphi_{\text{ext}}}{\partial n}\right)\bigg|_\Sigma = 8\pi(\mathbf{P}_S\mathbf{n})\big|_\Sigma, \quad \varphi(z=h) = 0$$

(1)

In order to apply all the following results to the anisotropic semiconductor one can make the substitution: $z \to z\sqrt{\varepsilon_a/\varepsilon_c}$ (e.g. $l \to l\sqrt{\varepsilon_a/\varepsilon_c}$), $\varepsilon \to \sqrt{\varepsilon_c\varepsilon_a}$, $R_d^2 = \dfrac{\varepsilon_a k_B T}{4\pi e^2 n_d}$. Here $\varepsilon_a$ and $\varepsilon_c$ are anisotropic dielectric permittivity values perpendicular and along the polar axis $z$.

Usually $\varepsilon \gg 1$, $R_d \sim (10^2 - 10^4)\,nm$ [10, 12] and $h \sim (10^3 - 10^5)\,nm$, so $\exp(-h/R_d) \ll 1$. Hereinafter we will use these assumptions. The solution of (1) can be found by means of the integral transformations, namely we obtained the potential in the form $\varphi(\mathbf{r}) = \varphi_q(\mathbf{r}) + \varphi_D(\mathbf{r})$, where:

**a)** The electric field potential $\varphi_q(\mathbf{r})$ is created by the point charge $q$ localized in air in the point $\mathbf{r}_0 = (0,0,-z_0)$. It is calculated in Appendix A (see A.7), namely in the case $\exp(-h/R_d) \ll 1$ we can use the approximation:

$$\varphi_q(\mathbf{r}) = \begin{cases} q\int_0^\infty dk J_0\left(k\sqrt{x^2+y^2}\right)\left(\exp(-k\cdot|z+z_0|) + \exp(k\cdot(z-z_0))\dfrac{k-\varepsilon\sqrt{k^2+R_d^{-2}}}{k+\varepsilon\sqrt{k^2+R_d^{-2}}}\right) & z<0 \\ 2q\int_0^\infty dk \dfrac{k\cdot J_0\left(k\sqrt{x^2+y^2}\right)}{k+\varepsilon\sqrt{k^2+R_d^{-2}}}\exp\left(-z\sqrt{k^2+R_d^{-2}}-kz_0\right) & z\geq 0 \end{cases} \quad (2)$$

**b)** The depolarization field potential $\varphi_D(\mathbf{r})$ is created by polarization reversal inside the semi-ellipsoidal domain with radius $d$ and length $l$. In this case $\varphi_D(\mathbf{r})$ is calculated in Appendix B (see (B.7)):

$$\varphi_D(\mathbf{r}) = \begin{cases} \varphi_{DE}(\mathbf{r}) + 8\pi P_S \int_0^\infty dk \dfrac{J_0\left(k\sqrt{x^2+y^2}\right)}{k+\varepsilon\sqrt{k^2+R_D^{-2}}}\exp\left(-\sqrt{k^2+R_D^{-2}}\cdot z\right)C_D(k), & z\geq 0 \\ 8\pi P_S \int_0^\infty dk \dfrac{J_0\left(k\sqrt{x^2+y^2}\right)}{k+\varepsilon\sqrt{k^2+R_D^{-2}}}\exp(k\cdot z)C_D(k), & z<0 \end{cases} \quad (3a)$$

Here $C_D(k) = -\sqrt{k^2+R_D^{-2}}\int_0^l dz' \exp\left(-\sqrt{k^2+R_D^{-2}}\cdot z'\right)d\sqrt{1-(z'/l)^2}\, J_1\left(kd\sqrt{1-(z'/l)^2}\right)$. The last term in (3a) is the surface potential created by domain butt. The first term $\varphi_{DE}(\mathbf{r})$ is the potential created by the polarization reversal inside the domain. It has the form:

$$\varphi_{DE}(\mathbf{r}) = -\dfrac{4\pi P_S}{\varepsilon}\int_0^\infty dk J_0\left(k\sqrt{x^2+y^2}\right)\int_0^l dz'\, d\sqrt{1-(z'/l)^2}\, J_1\left(kd\sqrt{1-(z'/l)^2}\right) \times \\ \times \left(sign(z-z')\exp\left(-\sqrt{k^2+R_d^{-2}}\cdot|z-z'|\right) + sign(z+z')\exp\left(-\sqrt{k^2+R_d^{-2}}\cdot|z+z'|\right)\right) \quad (3b)$$

In the case when screening radius $R_d$ is larger then curvature $r_C = d^2/l$ of the semi-ellipsoid apex, the potential $\varphi_{DE}(\mathbf{r})$ inside the prolate semi-ellipsoidal domain with $d \ll l$ acquires the form:

$$\varphi_{DE}(\mathbf{r}) \approx \dfrac{8\pi P_S}{\varepsilon}\dfrac{d^2}{l^2}\left(\ln\left(\dfrac{2l}{d}\right)-1\right)\exp\left(-\dfrac{l-\sqrt{z^2+l^2(x^2+y^2)/d^2}}{R_d}\right)\cdot z \quad (3c)$$

Note, that the surface $l - \sqrt{z^2 + l^2(x^2+y^2)/d^2} = R_d$ where depolarization field is mainly concentrated, corresponds to the "screening" ellipsoid $\dfrac{x^2+y^2}{d^2(1-R_d/l)^2} + \dfrac{z^2}{(l-R_d)^2} = 1$, with the same ellipticity $d/l$ as the domain one and semi-axes $(l - R_d)$, $d\sqrt{1 - R_d/l}$. The density of the screening charges $\rho_f \sim \varphi_{DE}(\mathbf{r})$ depends on the curvature of domain surface $\Sigma$. For example, the charge density accommodated near the semi-ellipsoid domain apex $z = l$ (where spontaneous polarization vector is normal to the domain surface) is maximal (see Fig.2). Surface potentials are usually neglected in papers (see e.g. [7]). To our mind, they could not be neglected in comparison with $\varphi_{DE}(\mathbf{r})$ at least near the sample surface, where $\varphi_{DE} = 0$.

The electrostatic energy of ferroelectrics is $\Phi_{el} = \dfrac{1}{8\pi}\int dv(\mathbf{D}\cdot\mathbf{E} - 4\pi\mathbf{P}_S\cdot\mathbf{E})$ (see chapter 2 in [13]). The excess of electrostatic energy $\Delta\Phi_{el}(d,l)$ caused by the origin of the semi-ellipsoidal domain with reversed polarization $-P_S \to +P_S$ is considered in details in Appendix C. Its general expression is:

$$\Delta\Phi_{el}(d,l) \approx \Delta\Phi_\rho(d,l) + \Delta\Phi_\sigma(d,l) + \frac{1}{2}q\varphi_D(r_0)$$
$$\Delta\Phi_\rho = -\frac{\varepsilon}{8\pi R_d^2}\int_{z>0} dv\left[(\varphi_q + \varphi_D)^2 - \varphi_q^2\right], \qquad (4)$$
$$\Delta\Phi_\sigma = \int_{\Sigma(z>0)} ds(\mathbf{P}_S\cdot\mathbf{n})(\varphi_q + \varphi_D) - \int_{\substack{(x^2+y^2)\le d^2 \\ z=0}} dxdy(\varphi_q + \varphi_D)P_S$$

The excess of electrostatic energy $\Delta\Phi_{el}(d,l) = \Phi_q(d,l) + \Phi_D(d,l)$ and domain wall surface energy $\Phi_C(d,l)$ contributes into the thermodynamic potential $\Phi(d,l)$:

$$\Phi(d,l) = \Phi_q(d,l) + \Phi_D(d,l) + \Phi_C(d,l). \qquad (5)$$

Below we consider every term in (5) with accuracy $O(\varepsilon^{-2})$ and $O(d^2/l^2)$.

**a)** Then excess of energy $\Phi_q(d,l)$ is caused by interaction between the AFM tip electric field and reversed polarization inside semi-ellipsoidal domain. It was calculated using the approximation (2) for $\varphi_q$. The Pade approximation for interaction energy $\Phi_q(d,l)$ over variable $R_d$ acquires the following form:

$$\Phi_q(d,l) = \frac{8\pi}{\varepsilon}P_S\, qR_d\, \frac{z_0 - \sqrt{z_0^2 + d^2}}{R_d + 2\sqrt{z_0^2 + d^2}} \qquad (6)$$

At $R_d \to \infty$ energy (6) coincides with the one calculated in [7].

**b)** The depolarization field energy $\Phi_D(d,l)$ is caused by polarization reversal within the semi-ellipsoidal domain surrounded by screening layer. The Pade approximation for depolarization field energy $\Phi_D(d,l)$ over variable $R_d$, acquires simple form only at $d \ll l$ and $\varepsilon \gg 1$, namely:

$$\Phi_D(d,l) = \frac{16\pi^2 P_S^2}{3\varepsilon} \frac{d^4 R_d (\ln(2l/d) - 1)}{l R_d + 4d^2 (\ln(2l/d) - 1)/3} + \frac{16\pi^2 P_S^2}{3\varepsilon} \frac{d^3 R_d}{\pi R_d/4 + 4d/3} \qquad (7)$$

At $d \ll R_d$ one obtains from (7), that $\Phi_D(d,l) \approx \Phi_{DE}(d,l) + \Phi_{DS}(d,l)$. The first term $\Phi_{DE}(d,l) \sim \frac{d^4}{l}$ is interaction energy of the real ($z = l$) and imaginary ($z = -l$) bound charges.

The last positive term $\Phi_{DS}(d,l) \approx \frac{64\pi P_S^2}{3\varepsilon} d^3$ is the intrinsic energy of the opposite polarized semi-ellipsoid butt [13], omitted in [7], [14], [15].

c) The correlation surface energy $\Phi_C(d,l)$ of the semi-ellipsoidal domain with $d \ll l$ has the form:

$$\Phi_C(d,l) = \pi d^2 \psi_S + \pi d l \psi_S \frac{1}{\sqrt{1 - (d/l)^2}} \arcsin\left(\sqrt{1 - (d/l)^2}\right) \approx \frac{\pi^2}{2} \psi_S d l \qquad (8)$$

We regard domain walls as infinitely thin, with homogeneous surface energy density $\psi_S$.

The obtained thermodynamic potential of the prolate semi-ellipsoidal domain with $d \ll l$ acquires the form:

$$\Phi(d,l) \approx \left( \begin{array}{c} \dfrac{8\pi}{\varepsilon} P_S q \dfrac{R_d \left(z_0 - \sqrt{z_0^2 + d^2}\right)}{R_d + 2\sqrt{z_0^2 + d^2}} + \dfrac{16\pi^2 P_S^2}{3\varepsilon} \dfrac{d^4 R_d (\ln(2l/d) - 1)}{l R_d + 4d^2 (\ln(2l/d) - 1)/3} + \\ + \dfrac{16\pi^2 P_S^2}{3\varepsilon} \dfrac{d^3 R_d}{\pi R_d/4 + 4d/3} + \dfrac{\pi^2}{2} \psi_S d l \end{array} \right) \qquad (9)$$

For the anisotropic ferroelectric-semiconductor (9) can be rewritten in the form:

$$\Phi(\delta,\lambda) = -W_q \cdot \frac{\rho\left(\sqrt{\delta^2 + \xi^2} - \xi\right)}{\rho + 2\sqrt{\delta^2 + \xi^2}} + W_C \cdot \delta\lambda + W_D \left( \frac{\rho\delta^4 (\ln(2\lambda/\delta) - 1)}{\lambda\rho + 4\delta^2 (\ln(2\lambda/\delta) - 1)/3} + \frac{\delta^3 \rho}{\pi\rho/4 + 4\delta/3} \right) \qquad (10)$$

Hereinafter the new variables have been introduced:

- dimensionless domain radius and length $\quad \delta = d/R_0, \quad \lambda = \sqrt{\varepsilon_a/\varepsilon_c}\,(l/R_0)$,
- the location of the effective charge $q$ $\quad \xi = z_0/R_0$,
- screening layer relative thickness $\quad \rho = R_d/R_0$,
- characteristic energies $\quad W_q = \dfrac{8\pi q P_S R_0}{\sqrt{\varepsilon_a \varepsilon_c}}, \quad W_C = \dfrac{\pi^2 R_0^2 \psi_S}{2}, \quad W_D = \dfrac{16\pi^2 P_S^2 R_0^3}{3\sqrt{\varepsilon_a \varepsilon_c}}.$

The equilibrium domain sizes $\{\lambda_e, \delta_e\}$ minimize the potential (10) and thus satisfy the system of equations:

$$\begin{cases} \dfrac{\partial \Phi(\delta,\lambda)}{\partial \lambda} = 0, & \dfrac{\partial \Phi(\delta,\lambda)}{\partial \delta} = 0, \\ \dfrac{\partial^2 \Phi(\delta,\lambda)}{\partial^2 \lambda} > 0, & \dfrac{\partial^2 \Phi(\delta,\lambda)}{\partial^2 \delta} > 0, \end{cases} \quad (11)$$

Simple approximate formula for the dependence $\lambda_e(\delta_e)$ could be obtained from (10)-(11), namely from the equations

$$0 = \frac{\partial \Phi(\delta,\lambda)}{\partial \lambda} = W_C \cdot \delta - W_D \frac{\delta^4 (\ln(2\lambda/\delta) - 2)\rho^2}{(\lambda\rho + 4\delta^2 (\ln(2\lambda/\delta)-1)/3)^2};$$

$$0 = \frac{\partial \Phi(\delta,\lambda)}{\partial \delta} = -W_q \cdot \frac{\delta\rho(\rho + 2\xi)}{\sqrt{\delta^2 + \xi^2}(\rho + 2\sqrt{\delta^2 + \xi^2})^2} + W_C \cdot \lambda + \quad (12)$$

$$+ W_D \left( \frac{3\delta\rho}{2} - \frac{\lambda\delta\rho^2(2\delta^2 + 3\lambda\rho)}{2(\lambda\rho + 4\delta^2(\ln(2\lambda/\delta)-1)/3)^2} + \frac{3\pi\delta^2\rho^2/4 + 8\delta^3\rho/3}{(\pi\rho/4 + 4\delta/3)^2} \right).$$

For the usual case $3\lambda_e \rho \gg 4\delta_e^2$ and $W_C \ll W_D$ we found the following approximate equations:

$$\lambda_e = \delta_e^{3/2} \sqrt{\frac{W_D}{W_C}(\ln(2\lambda_e/\delta_e) - 2)} \approx \delta_e^{3/2} \sqrt{\frac{W_D}{W_C}\left(\ln\left(2\sqrt{\frac{W_D}{W_C}}\delta_e\right) - 2\right)} \quad (13a)$$

$$W_q \approx \left( W_D \left( \frac{3\pi\delta_e\rho/4 + 8\delta_e^2/3}{(\pi\rho/4 + 4\delta_e/3)^2} - \frac{\delta_e^2}{\lambda_e\rho} \right) + W_C \frac{\lambda_e}{\delta_e\rho} \right) \cdot \frac{\sqrt{\delta_e^2 + \xi^2}}{(\rho + 2\xi)} \left(\rho + 2\sqrt{\delta_e^2 + \xi^2}\right)^2 \quad (13b)$$

Usually $W_C \ll W_D$ and the second term in (13b) is negligibly small. In this case equation (13b) for $\delta_e$ can be easily solved by graphic method, namely the dependence of $\delta_e$ over $W_q$ can be plotted. Keeping in mind, that $W_q$ is proportional to AFM tip charge $q \approx UC_t$ ($U$ is applied voltage pulse, $C_t$ is the effective capacity of the system AFM apex-air-semiconductor-bottom electrode), we obtain the dependences $\delta_e(U)$ and $\lambda_e(U)$.

By definition effective capacity $C_t$ of the system "metal AFM tip-semiconductor" could be estimated as the ratio of the tip charge $q$ to the potential $\varphi_q$ at the tip apex (see (2) and Fig.2), namely at $\varepsilon \gg 1$ and $R_0 \gg \Delta R$:

$$C_t = \frac{q}{\varphi_q(x=0, y=0, z=-\Delta R)} \sim \left( \frac{1}{z_0 - \Delta R} - \frac{1}{z_0 + \Delta R}\left(1 - \frac{2R_d}{\varepsilon(R_d + z_0 + \Delta R)}\right) \right)^{-1} \sim \frac{z_0^2}{2\Delta R} \quad (14)$$

In the zero approximation of the spherical model $z_0 \approx (R_0 + \Delta R)$ (see [7] and references therein). Surely (14) is approximate mainly due to the fact, that the charged tip potential near the sample surface was calculated as the point charge one. The validity of such assumptions for dielectric sample and more sophisticated models are discussed in [11]. Our over-simplified result (14) obtained for semiconductor sample is the self-consistent zero approximation. To our mind, both the potential $\varphi_q$ and the capacity $C_t$ must be improved simultaneously in the case of inadequate agreement with experimental data. In the next section we present the dependences $\delta_e(U)$ and $\lambda_e(U)$, then compare our results with experiments.

### 3. EQUILIBRIUM DOMAIN SIZES: CALCULATIONS AND COMPARISON WITH EXPERIMENTS

The dependences (13) of equilibrium domain radius $\delta_e = d_e/R_0$ and length $\lambda_e = d_e/R_0$ over applied voltage $U$ (in $kV$) for different screening radiuses $\rho = R_d/R_0$ are shown in the Figs 3. The chosen characteristic energies values are typical for such ilmenites as LiNbO$_3$ and LiTaO$_3$ single crystals with $W_D \sim 10^{-6}\,Erg$, $W_C \sim 10^{-9}\,Erg$, $W_q(U) \sim U \cdot 10^{-3}\,Erg$.

We would like to underline, that screening carriers not only decrease depolarization field inside the domain, but also they shield the AFM tip electric field inside the sample. As a result, screening radius decrease leads to the essential decrease of the equilibrium domain sizes (see the lowest curves in the Figs 3). Depolarization field caused by the uncompensated surface charges at the domain butt $z=0$ also leads to the essential decrease of the of the domain sizes (compare dashed and solid curves). Note that this field did not appear in the system considered by Landauer [6] due to the complete screening of surface bond charges by the free charge inside the upper electrode.

Let us discuss the question of domain stability after applied voltage is turned off. In the majority of experiments [1], [2], [15] reversed domains remains their initial shape and sizes during many days and weeks. This fact is extremely useful for the applications and has the following explanation within the framework of the proposed model. In the case $W_C << W_D$ the domain radius is determined from the balance between the negative interaction energy $\Phi_q(d,l)$ and the positive domain butt depolarization energy $\Phi_{DS}(d,l)$ (see (13b)). When voltage $U$ is turned off the interaction field disappears as proportional to $U$, the domain butt depolarization field vanishes due to the sluggish recharging of the surface traps in the absence of the AFM tip electric field. Thus the total energy of the system remained practically constant and the domain is in the equilibrium. Note, that the stabilization process could be related to the pinning of the domain wall [16], but the domain kinetics is out of the scope of our paper.

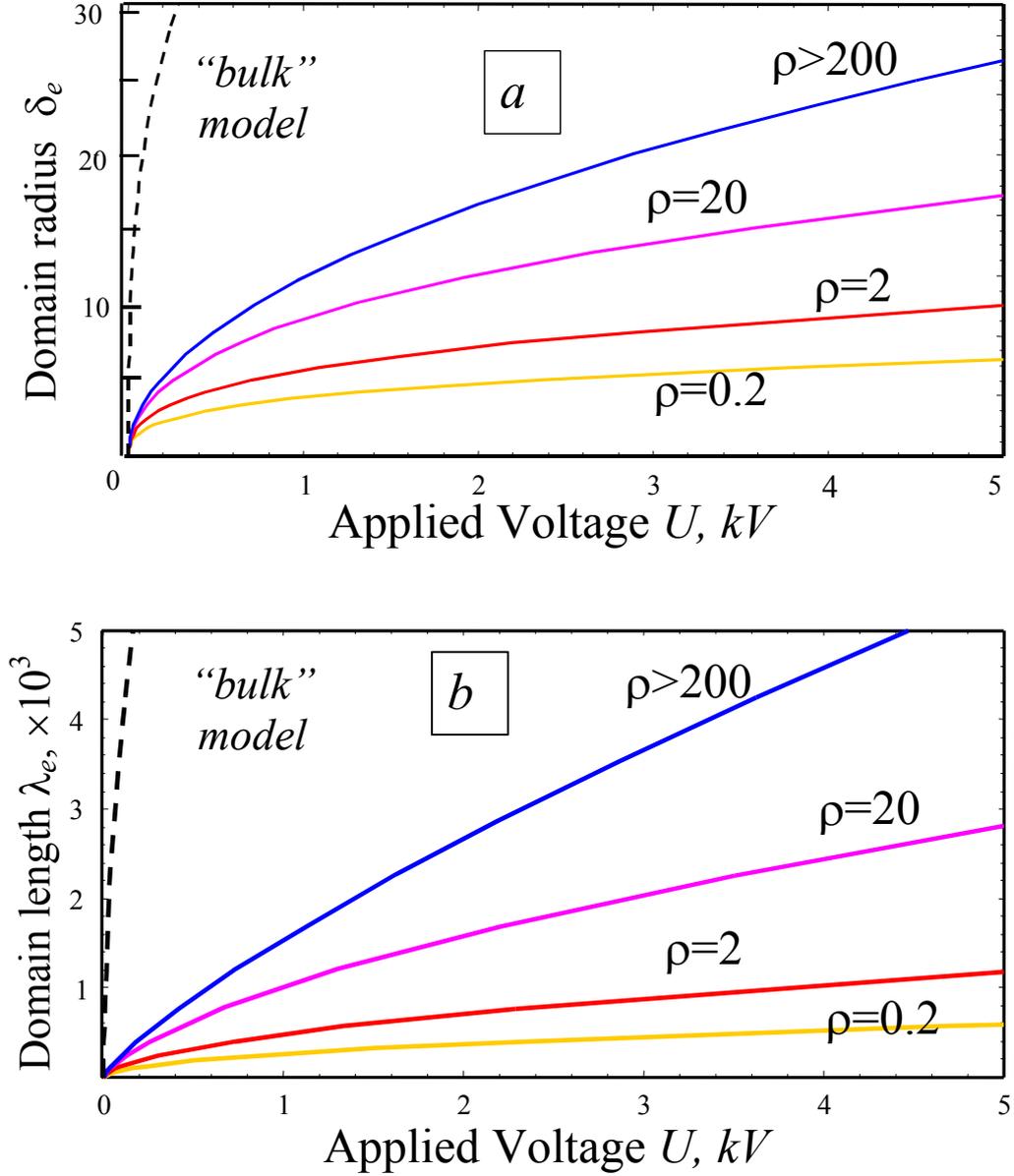

**Figure 3.** The dependences of equilibrium domain radius $\delta_e = d_e/R_0$ (a) and length $\lambda_e = d_e/R_0$ (b) over applied voltage $U$ at different screening radius $\rho = R_d/R_0$ and $\xi = 1$. Characteristic energies values $W_D \sim 3 \cdot 10^{-6} \, Erg$, $W_C \sim 5 \cdot 10^{-9} \, Erg$, $W_q(U) \sim 5 \cdot 10^{-3} \, Erg \cdot U$ ($U$ in $kV$).

Now let us apply our theoretical results to the micro-domain formation in LiNbO$_3$ single crystals using high-voltage AFM. In experiments [1], [15] AFM tip radius was $R_0 = 50 nm$, distance $\Delta R \sim 0.1 nm$, maximum applied voltage pulse value $U_{max} = 4 kV$ with duration up to 5min., sample thickness $h = 0.15mm, 1mm$. For LiNbO$_3$ at room temperature $\varepsilon_a = 84$, $\varepsilon_c = 30$, $\psi_S \approx 50\,mJ/m^2 = 50\,Erg/cm^2$, $P_S \approx 50\mu C/cm^2 \approx 1.5\cdot10^5\,Q_{CGSE}/cm^2$, so we estimated from (14) the system capacity $C_t \sim 1.26\cdot10^{-3}\,cm$ in CGSE units. The comparison of experimental results and our calculations is presented in the Fig.4.

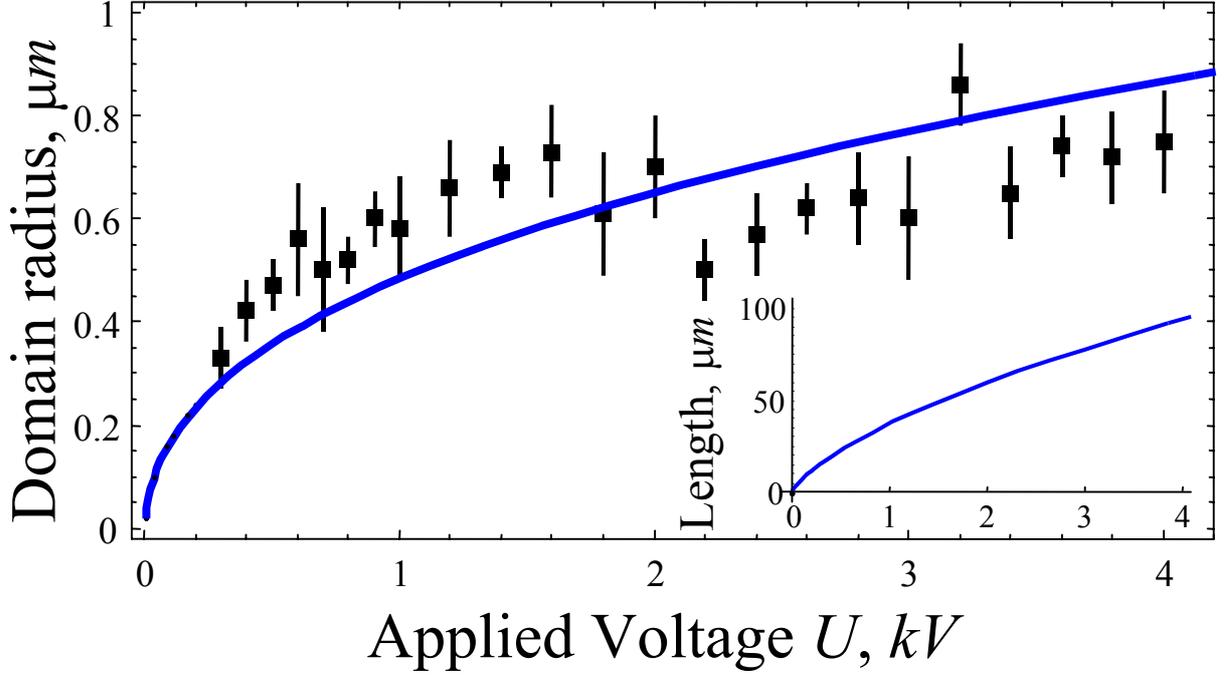

**Figure 4.** The dependences of equilibrium domain radius and length in LiNbO$_3$ over applied voltage $U$. Squares with error bars are experimental data from [1] for domain radius $d$, domain length was estimated as 150-250$\mu m$ [15]. Solid curve is our fitting at $R_d = 2\mu m$, $\Delta R = 0.1nm$, $z_0 = 50.1nm$ and characteristic energies $W_q(U_{max}) = 6.28\cdot10^{-3}\,Erg$, $W_D = 2.95\cdot10^{-6}\,Erg$, $W_C = 6.17\cdot10^{-9}\,Erg$.

The obtained fitting value of Debye screening radius $R_d = 2\mu m$ is in reasonable agreement with the estimations $R_d \sim (10^2 - 10^4)nm$ valid for the most ferroelectrics-semiconductors with unavoidable growth defects [12]. Though our modeling and experimental points are in a relatively good agreement, several notes should be made, namely at small distances $\Delta R < 1nm$ the AFM tip could touch the surface roughness, so the electroelastic coupling may be significant in the case, also the tip capacity $C_t$ cannot be directly calculated together with $z_0$, they could be treated as fitting parameters varied in the range determined by (14).

Let us apply our theoretical results to the nano-domain formation in congruent LiTaO$_3$ single crystals using high-voltage AFM. In experiment [2] AFM tip radius was $R_0 = 25 nm$, maximum applied voltage pulse value $U_{max} = 11 V$ with duration up to $10^{-4}$ s, and LiTaO$_3$ thin films have thickness $h = 55 \div 83\, nm$. For LiTaO$_3$ at $T$=300K $\varepsilon_a = 51$, $\varepsilon_c = 45$, $P_S \approx 50 \mu C/cm^2$, $\psi_S \approx 40 mJ/m^2$, so we estimated from (14) the system capacity in CGSE units, namely $C_t \approx 3.15 \cdot 10^{-4} cm$. The comparison of experimental results and our calculations is presented in the Fig.5.

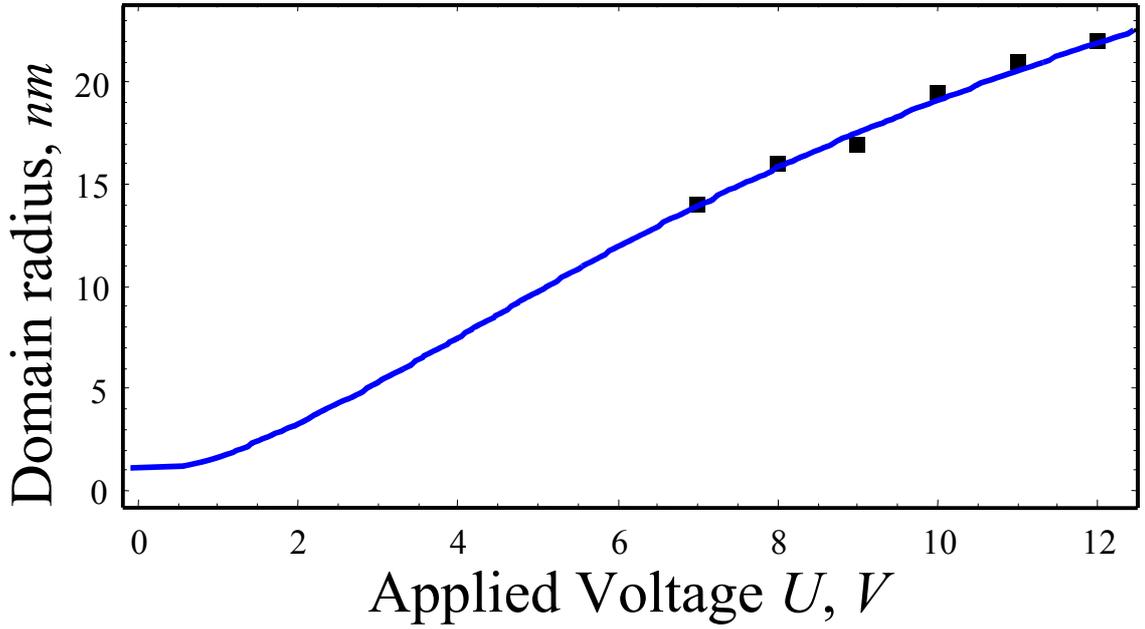

**Figure 5.** The dependences of domain radius in LiTaO$_3$ over applied voltage $U$. Domains penetrates through the sample. Squares are experimental data from [2] for domain radius $d$. Solid curve is our fitting at $R_d = 30\, nm$, $\Delta R \approx 0.1\, nm$, $z_0 = 25.1\, nm$ and characteristic energies $W_q(U_{max}) = 2.27 \cdot 10^{-6}\, Erg$, $W_D = 3.86 \cdot 10^{-7}\, Erg$, $W_C = 1.23 \cdot 10^{-9}\, Erg$.

The relatively small fitting value of Debye screening radius $R_d = 30\, nm$ could be explained by the presence of numerous lithium vacancies reported by authors [2] as well as by the carriers injection from the tip caused by the strong electric field inside the thin film. We calculated that $\exp(-h/R_d) \approx 0.1$, and so the main assumption of our theory $\exp(-h/R_d) \ll 1$ is valid, but the question about the applicability of our "equilibrium" calculations remains open due to small duration of applied voltage pulses. Despite this warning, we obtained rather good quantitative agreement between our modeling and experimental points.

## CONCLUSION

- The equilibrium shape of micro- and nano-domains caused by electric field of atomic force microscope tip in ferroelectric semiconductor crystals has been calculated. For the first time we modified the Landauer model allowing for semiconductor properties of the sample and the surface energy of the domain butt. Additional surface energy of the domain butt and "screening" of the depolarization and interaction energies lead to the essential decrease of the equilibrium domain sizes.

- We expressed the radius and length of the domain though the material screening radius, spontaneous polarization value, dielectric permittivity tensor, atomic force microscope tip charge $q \sim U$, radius of curvature and distance from the sample surface. The obtained dependence of domain radius via applied voltage $U$ is in a good quantitative agreement with the ones of submicron ferroelectric domains recorded by high-voltage atomic force and scanning probe microscopy in $LiNbO_3$ and $LiTaO_3$ crystals, in contrast to the over-estimated dependencies calculated within Landauer model evolved for perfect polar dielectrics covered with superconducting electrodes.

- Therefore we hope, that our results will help one to determine the necessary recording conditions and appropriate ferroelectric medium in order to obtain the stable domains with minimum lateral size.

## ACKNOWLEDGMENTS

Authors are grateful to Profs. N.V. Morozovsky and S.L. Bravina for useful discussions of our model.

## REFERENCES.

## APPENDIX A

For the extrinsic semiconductor with donors concentration $n_d$, the free charges bulk density $\rho_f$ is determined via electric field potential $\varphi(\mathbf{r})$, as follows:

$$\rho_f(\mathbf{r}) = e\left(N n_d \exp\left(-\frac{Ne\varphi(\mathbf{r})}{k_B T}\right) - n_0 \exp\left(\frac{e\varphi(\mathbf{r})}{k_B T}\right)\right) \qquad (A1)$$

Hereinafter we consider the case $|Ze\varphi(\mathbf{r})/k_B T| \ll 1$. Taking into account the electro neutrality condition $Zn_d = n_0$, one can find that

$$\rho_f(\mathbf{r}) \approx -\frac{\varepsilon \varphi(\mathbf{r})}{4\pi R_d^2}, \quad R_d^2 = \frac{\varepsilon k_B T}{4\pi e^2 (Z^2 n_d + n_0)}. \qquad (A.2)$$

The external electric field potential $\varphi_q(\mathbf{r})$ created by the point charge $q$ localized in air in the point $r_0 = (0,0,-z_0)$, inside the film $0 \le z \le h$ filled by isotropic semiconductor with dielectric permittivity $\varepsilon$ could be found from the boundary problem:

$$\begin{aligned}\Delta\varphi_0(\mathbf{r}) &= -4\pi q \delta(x,y,z+z_0), \qquad z \le 0, \\ \Delta\varphi_q(\mathbf{r}) - \frac{\varphi_q(\mathbf{r})}{R_d^2} &= 0, \qquad 0 \le z \le h, \\ \varphi_0(z=0) = \varphi_q(z=0), \quad \left(\frac{\partial \varphi_0}{\partial z} - \varepsilon \frac{\partial \varphi_q}{\partial z}\right)\bigg|_{z=0} &= 0, \quad \varphi_q(z=h) = 0\end{aligned} \qquad (A.3)$$

The general solution of (A.3) could be found in the form:

$$\varphi_0(\mathbf{r}) = \int_0^\infty dk J_0\!\left(k\sqrt{x^2+y^2}\right)\left(\exp(-k\cdot|z+z_0|) + q\cdot\exp(k\cdot z) C_0(k)\right)$$

$$\varphi_q(\mathbf{r}) = \int_0^\infty dk J_0\!\left(k\sqrt{x^2+y^2}\right)\left(\exp\!\left(-\sqrt{k^2+R_d^{-2}}\cdot z\right) C_q(k) + \exp\!\left(\sqrt{k^2+R_d^{-2}}\cdot z\right) B_q(k)\right)$$ (A.4)

Hereinafter $J_0$ is Bessel function of zero order. Functions $C_0$, $B_q$ and $C_q$ can be determined from the boundary conditions. Finally we obtained that:

$$\varphi_0(\mathbf{r}) = q\int_0^\infty dk J_0\!\left(k\sqrt{x^2+y^2}\right)\left(\begin{array}{c}\exp(-k\cdot|z+z_0|) + \exp(k\cdot(z-z_0))\times \\ \dfrac{k\!\left(1-\exp\!\left(-2h\sqrt{k^2+R_d^{-2}}\right)\right) - \varepsilon\sqrt{k^2+R_d^{-2}}\!\left(1+\exp\!\left(-2h\sqrt{k^2+R_d^{-2}}\right)\right)}{k\!\left(1-\exp\!\left(-2h\sqrt{k^2+R_d^{-2}}\right)\right) + \varepsilon\sqrt{k^2+R_d^{-2}}\!\left(1+\exp\!\left(-2h\sqrt{k^2+R_d^{-2}}\right)\right)}\end{array}\right)$$

$$\varphi_q(\mathbf{r}) = q\int_0^\infty dk J_0\!\left(k\sqrt{x^2+y^2}\right)\dfrac{2k\!\left(\exp\!\left(-z\sqrt{k^2+R_d^{-2}} - k z_0\right) - \exp\!\left(-(2h-z)\sqrt{k^2+R_d^{-2}} - k z_0\right)\right)}{k\!\left(1-\exp\!\left(-2h\sqrt{k^2+R_d^{-2}}\right)\right) + \varepsilon\sqrt{k^2+R_d^{-2}}\!\left(1+\exp\!\left(-2h\sqrt{k^2+R_d^{-2}}\right)\right)}$$

(A.5)

Usually $\varepsilon \gg 1$, $R_d \sim (10-10^2)\,nm$ and $h \sim (10^2-10^4)\,nm$, so $\exp(-h/R_d) \ll 1$, thus using expansion

$$\dfrac{\exp\!\left[-\sqrt{x^2+y^2+z^2}/R_d\right]}{\sqrt{x^2+y^2+z^2}} = \int_0^\infty dk \dfrac{k J_0\!\left(k\sqrt{x^2+y^2}\right)}{\sqrt{k^2+R_d^{-2}}}\exp\!\left(-\sqrt{k^2+R_d^{-2}}\cdot|z|\right),$$ (A.6)

one can obtain from (A.5) well-known potentials at $\varepsilon \gg 1$ and $\exp(-h/R_d) \ll 1$:

$$\varphi_0(\mathbf{r}) \approx q\int_0^\infty dk J_0\!\left(k\sqrt{x^2+y^2}\right)\left(\exp(-k\cdot|z+z_0|) + \exp(k\cdot(z-z_0))\dfrac{k-\varepsilon\sqrt{k^2+R_d^{-2}}}{k+\varepsilon\sqrt{k^2+R_d^{-2}}}\right) \approx$$

$$\approx \dfrac{q}{\sqrt{x^2+y^2+(z_0+z)^2}} - q\dfrac{\varepsilon-1}{\varepsilon+1}\dfrac{1}{\sqrt{x^2+y^2+(z_0-z)^2}}$$ (A.7a)

Inside the sample:

$$\varphi_q(\mathbf{r}) \approx 2q\int_0^\infty dk \dfrac{k\cdot J_0\!\left(k\sqrt{x^2+y^2}\right)}{k+\varepsilon\sqrt{k^2+R_d^{-2}}}\exp\!\left(-z\sqrt{k^2+R_d^{-2}} - k z_0\right) \le \dfrac{2q}{\varepsilon+1}\dfrac{\exp(-z/R_d)}{\sqrt{x^2+y^2+(z_0+z)^2}}$$ (A.7b)

In order to apply this result to the anisotropic semiconductor one can use the substitution:

$$z \to z/\sqrt{\varepsilon_c/\varepsilon_a},\quad \varepsilon \to \sqrt{\varepsilon_c\varepsilon_a},\quad R_d^2 = \dfrac{\varepsilon_a k_B T}{4\pi e^2\!\left(Z^2 n_d + n_0\right)}.$$

## APPENDIX B

The depolarization field potential $\varphi_D(\mathbf{r})$ satisfies the Poisson equation with the interfacial conditions $D_{n\,\text{int}} = D_{n\,\text{ext}}$ on the domain surface $\Sigma$ and $D_{n\,\text{ext}} - D_{n\,\text{int}} = 4\pi\sigma_S$ at the surface $z=0$. It can be written as the boundary problem:

$$\Delta\varphi_{D0}(\mathbf{r}) = 0, \quad z \leq 0,$$

$$\varphi_{D0}(z=0) = \varphi_D(z=0), \quad \left(\frac{\partial\varphi_{D0}}{\partial z} - \varepsilon\frac{\partial\varphi_D}{\partial z}\right)\bigg|_{z=0} = \begin{cases} -8\pi P_S, & \sqrt{x^2+y^2} < d \\ 0, & \sqrt{x^2+y^2} > d \end{cases}$$

$$\Delta\varphi_D(\mathbf{r}) - \frac{\varphi_D(\mathbf{r})}{R_d^2} = 0, \quad 0 \leq z \leq h \qquad (B.1)$$

$$\varepsilon\left(\frac{\partial\varphi_{D\,\text{int}}}{\partial n} - \frac{\partial\varphi_{D\,\text{ext}}}{\partial n}\right)\bigg|_{\Sigma} = 8\pi(\mathbf{P}_S\,\mathbf{n})\big|_{\Sigma}, \quad \varphi_D(z=h) = 0$$

Firstly let us calculate the part $\varphi_{DE}(\mathbf{r})$ of potential $\varphi_D(\mathbf{r}) = \varphi_{DE}(\mathbf{r}) + \varphi_{DS}(\mathbf{r})$ created by polarized ellipsoid in the infinite isotropic semiconductor with permittivity $\varepsilon$:

$$\Delta\varphi_{DE}(\mathbf{r}) - \frac{\varphi_{DE}(\mathbf{r})}{R_d^2} = 0,$$

$$\varepsilon\left(\frac{\partial\varphi_{DE\,\text{int}}}{\partial n} - \frac{\partial\varphi_{DE\,\text{ext}}}{\partial n}\right)\bigg|_{\Sigma} = 8\pi(\mathbf{P}_S\,\mathbf{n})\big|_{\Sigma}, \qquad (B.2)$$

$$\varphi_{DE}(z=0) = 0$$

The solution of (B.2) can be found by means of the Green function method and integral transformations, namely we obtained:

$$\varphi_{DE}(\mathbf{r}) = \int_\Sigma ds' \frac{2(\mathbf{P}_S\,\mathbf{n}(\mathbf{r}'))}{\varepsilon} \cdot \frac{\exp(-|\mathbf{r}-\mathbf{r}'|/R_d)}{4\pi|\mathbf{r}-\mathbf{r}'|} =$$

$$= -\frac{4\pi P_S}{\varepsilon}\int_0^\infty dk\, J_0\!\left(k\sqrt{x^2+y^2}\right)\int_0^l dz'\,\frac{d}{dk}J_0\!\left(kd\sqrt{1-(z'/l)^2}\right)\times \qquad (B.3)$$

$$\times\left(sign(z-z')\exp\!\left(-\sqrt{k^2+R_d^{-2}}\cdot|z-z'|\right) + sign(z+z')\exp\!\left(-\sqrt{k^2+R_d^{-2}}\cdot|z+z'|\right)\right)$$

Here we integrate over the whole ellipsoid in order to satisfy the condition $\varphi_{DE}(z=0) = 0$ and then to continue $\varphi_{DE}(z \leq 0) = 0$. One can see that for perfect dielectric $R_d \to \infty$, and thus the solution $\varphi_{DEP}(\mathbf{r}) = \int_\Sigma ds' \frac{2(\mathbf{P}_S\,\mathbf{n})}{4\pi\varepsilon|\mathbf{r}-\mathbf{r}'|}$ of this problem without screening is the known Coulomb potential created by bound surface charge with density $\sigma_b(\mathbf{r}) = 2(\mathbf{P}_S\,\mathbf{n}(\mathbf{r}))$ and calculated in [13] with the help of ellipsoidal coordinates. In contrast, for ideal conductor $R_d \to 0$, and the solution $\varphi_D(\mathbf{r}) \to 0$ as it should be expected inside the metal.

Allowing for $\sigma_b(\mathbf{r}) \geq 0$, it is easy to obtain from (B.3) the following estimation for potential $\varphi_{DE}(\mathbf{r})$:

$$0 \leq \varphi_{DE}(\mathbf{r}) \leq \varphi_{DEP}(\mathbf{r}) \exp\left(-\frac{r_\Sigma(\mathbf{r})}{R_d}\right). \tag{B.4}$$

Hereinafter $r_\Sigma(\mathbf{r})$ is the distance between the point $\mathbf{r}$ and the domain boundary $\Sigma(d,l)$. It is clear from (B.4), that for small enough screening radius $R_d$ potential $\varphi_{DE}(\mathbf{r})$ is concentrated inside the layer $r_\Sigma(\mathbf{r}) \leq R_d$. Keeping in mind exact expression for $\varphi_{DEP}(\mathbf{r})$ and (B.4), we obtained the relatively simple approximation for $\varphi_{DE}(\mathbf{r})$ for elongated domain with $l \gg d$ and $R_d \geq d^2/l$:

$$\varphi_{DE}(\mathbf{r}) \leq \begin{cases} \dfrac{8\pi P_S}{\varepsilon} \dfrac{d^2}{l^2}\left(arcth\left(\sqrt{1-\dfrac{d^2}{l^2}}\right)-1\right) \exp\left(-\dfrac{l-\sqrt{z^2+l^2(x^2+y^2)/d^2}}{R_d}\right) \cdot z, & s \leq 0 \\ \dfrac{8\pi P_S}{\varepsilon} \dfrac{d^2}{l^2}\left(arcth\left(\sqrt{\dfrac{l^2-d^2}{s+l^2}}\right)-\dfrac{l}{\sqrt{s+l^2}}\right) \exp\left(-\dfrac{\sqrt{z^2+l^2(x^2+y^2)/d^2}-l}{R_d}\right) \cdot z, & s \geq 0 \end{cases}$$

$$\tag{B.5}$$

Here $s(x,y,z)$ is the one of ellipsoidal coordinates $\dfrac{x^2+y^2}{d^2+s}+\dfrac{z^2}{l^2+s}=1$ ($s=0$ corresponds to the boundary of domain).

Now let us calculate the surface screening potential $\varphi_{DS}(\mathbf{r})$ created by ferroelectric-semiconductor domain butt.

$$\begin{aligned} \Delta\varphi_{D0}(\mathbf{r}) &= 0, \quad z \leq 0, \\ \varphi_{D0}(z=0) &= \varphi_{DS}(z=0), \\ \left(\dfrac{\partial \varphi_{D0}}{\partial z} - \varepsilon\dfrac{\partial(\varphi_{DS}+\varphi_{DE})}{\partial z}\right)\bigg|_{z=0} &= \begin{cases} -8\pi P_S, & \sqrt{x^2+y^2} < d \\ 0, & \sqrt{x^2+y^2} > d \end{cases} \\ \Delta\varphi_{DS}(\mathbf{r}) - \dfrac{\varphi_{DS}(\mathbf{r})}{R_d^2} &= 0, \quad 0 \leq z \end{aligned} \tag{B.6}$$

The solution of (B.6) can be found using expansions from Appendix A. It is easy to obtain from (B.3) that:

$$\varphi_{D0}(\mathbf{r}) = 8\pi P_S \int_0^\infty dk \frac{J_0\left(k\sqrt{x^2+y^2}\right)}{k+\varepsilon\sqrt{k^2+R_D^{-2}}} \exp(k\cdot z) C_D(k)$$

$$\varphi_{DS}(\mathbf{r}) = 8\pi P_S \int_0^\infty dk \frac{J_0\left(k\sqrt{x^2+y^2}\right)}{k+\varepsilon\sqrt{k^2+R_D^{-2}}} \exp\left(-z\cdot\sqrt{k^2+R_D^{-2}}\right) C_D(k)$$

(B.7)

hereinafter $C_D(k) = \sqrt{k^2+R_D^{-2}} \int_0^l dz' \exp\left(-\sqrt{k^2+R_D^{-2}}\cdot z'\right) \frac{d}{dk} J_0\left(kd\sqrt{1-(z'/l)^2}\right)$.

In the case $d \ll l$ derivative $\partial\varphi_{DE}/\partial z$ in the boundary conditions of (B.6) can be approximated as:

$$\frac{\partial\varphi_{DE}(z=0)}{\partial z} \approx \frac{8\pi P_S}{\varepsilon} d \int_0^\infty dk J_0\left(k\sqrt{x^2+y^2}\right) J_1(kd) \exp\left(-\sqrt{k^2+R_D^{-2}}\cdot l\right) + O(d^2/l^2) \quad (B.8)$$

So, the solution (B.7) can be approximated as:

$$\varphi_{D0}(\mathbf{r}) \approx -8\pi P_S d \int_0^\infty dk \frac{J_0\left(k\sqrt{x^2+y^2}\right)}{k+\varepsilon\sqrt{k^2+R_d^{-2}}} J_1(kd)\cdot\exp(k\cdot z)\cdot\left[1-\exp\left(-l\cdot\sqrt{k^2+R_d^{-2}}\right)\right],$$

(B.9a)

$$\varphi_{DS}(\mathbf{r}) \approx -8\pi P_S d \int_0^\infty dk \frac{J_0\left(k\sqrt{x^2+y^2}\right)}{k+\varepsilon\sqrt{k^2+R_d^{-2}}} J_1(kd)\cdot\exp\left(-z\cdot\sqrt{k^2+R_d^{-2}}\right)\cdot\left[1-\exp\left(-l\cdot\sqrt{k^2+R_d^{-2}}\right)\right]$$

(B.9b)

At $\varepsilon \gg 1$ and $d \ll l$ we obtain from (B.5) and (B.9b) the following approximations for $\varphi_D$ valid (compare with (A.7a):

$$\varphi_{DS}(\mathbf{r}) \sim -4\pi P_S d^2 \int_0^\infty dk \frac{kJ_0\left(k\sqrt{x^2+y^2}\right)}{k+\varepsilon\sqrt{k^2+R_d^{-2}}} \exp\left(-z\cdot\sqrt{k^2+R_d^{-2}}\right) \quad (B.10)$$

## APPENDIX C

The electrostatic energy is created by the surface charges $\sigma_b(\mathbf{r})$ located on the domain surfaces $\Sigma$, $\sigma_b$ and $\sigma_S$ at $z=0$ related to the spontaneous polarization discontinuity, as well as by the bulk charges $\rho_f(\mathbf{r}) = -\varepsilon\varphi(\mathbf{r})/4\pi R_d^2$. It has the form:

$$\Phi_{el}(\varphi_q,\varphi_D) = \frac{1}{8\pi}\int_V dv(\mathbf{D}\cdot\mathbf{E}-4\pi\mathbf{P}_S\cdot\mathbf{E}) \equiv \Phi_\rho(\varphi_q,\varphi_D) + \Phi_\sigma(\varphi_q,\varphi_D) + \frac{1}{2}q\varphi_{D0}(r_0)$$

$$\Phi_\rho(\varphi_q,\varphi_D) = \frac{1}{2}\int_V dv\,\rho_f(\mathbf{r})\varphi(\mathbf{r}) = -\frac{\varepsilon}{8\pi R_d^2}\int_{z>0} dv(\varphi_q(\mathbf{r})+\varphi_D(\mathbf{r}))^2, \quad (C.1)$$

$$\Phi_\sigma = \frac{1}{2}\int_S ds\,\sigma_b\,\varphi = \int_\Sigma ds(\mathbf{P}_S\cdot\mathbf{n})(\varphi_q+\varphi_D) - \int_{\substack{(x^2+y^2)\leq d^2 \\ z=0}} dxdy(\varphi_q+\varphi_D)P_S$$

Note, that the first term in $\Phi_\sigma$ acquires the form similar to the one considered in [6, 7] in accordance with Gauss theorem. Let us find the excess of electrostatic energy $\Delta\Phi_{el} = \Phi_{el}(\varphi_q,\varphi_D) - \Phi_{el}(\varphi_q,0)$ caused by polarization reversal inside the domain $-P_S \to +P_S$:

$$\Delta\Phi_{el} \approx \Delta\Phi_\rho + \Delta\Phi_\sigma + \frac{1}{2}q\varphi_{D0}(r_0)$$

$$\Delta\Phi_\rho = -\frac{\varepsilon}{8\pi R_d^2}\int_{z>0} dv\left[(\varphi_q(\mathbf{r})+\varphi_D(\mathbf{r}))^2 - \varphi_q(\mathbf{r})^2\right], \quad (C.2)$$

$$\Delta\Phi_\sigma = \int_{\Sigma(z>0)} ds(\mathbf{P}_S\cdot\mathbf{n})(\varphi_q+\varphi_D) - \int_{\substack{(x^2+y^2)\leq d^2 \\ z=0}} dxdy(\varphi_q+\varphi_D)P_S$$

In the case $\exp(-h/R_D)\ll 1$, $d\ll l$ and $\varepsilon\gg 1$, up to the terms proportional to $\varepsilon^{-2}$ the excess of electrostatic energy acquires the form $\Delta\Phi_{el} = \Phi_D(d,l) + \Phi_q(d,l)$.

1) The interaction energy $\Phi_q(d,l)$ between the domain and the AFM tip acquires the following form (see (A.7) and (B.7)):

$$\Phi_q(d,l) = \int_{\Sigma(z>0)} ds(\mathbf{P}_S\cdot\mathbf{n})\varphi_q - \int_{\substack{(x^2+y^2)\leq d^2 \\ z=0}} dxdy\,\varphi_q P_S - \frac{\varepsilon}{4\pi R_d^2}\int_{z>0} dv\,\varphi_D\,\varphi_q + \frac{q}{2}\varphi_{D0}(r_0) \approx$$

$$\approx \begin{cases} \dfrac{8\pi}{\varepsilon}P_S qR_d\dfrac{(z_0-\sqrt{z_0^2+d^2})}{2\sqrt{z_0^2+d^2}} + O\left(\left(\dfrac{d}{l}\right)^2 \exp(-l/R_d)\right), & R_d\to 0 \\[2ex] \dfrac{8\pi}{\varepsilon}P_S q(z_0-\sqrt{z_0^2+d^2}) + O\left(\left(\dfrac{d}{l}\right)^2 \exp(-l/R_d)\right), & R_d\to\infty \end{cases}$$

The [1/1] Pade approximation for interaction energy $\Phi_q(d,l)$ over variable $R_d$, which has 5% accuracy, acquires the following form:

$$\Phi_q(d,l) = \frac{8\pi P_S\,q}{\varepsilon}\frac{R_d(z_0-\sqrt{z_0^2+d^2})}{R_d+2\sqrt{z_0^2+d^2}} \quad (C.3)$$

2) The depolarization field energy related to the surfaces $z=0$ and $\Sigma$ has the following form:

$$\Phi_D(d,l) = \Phi_{DV}(d,l) + \Phi_{DS}(d,l) \tag{C.4}$$

The "bulk" depolarization field energy $\Phi_{DV}(d,l)$ has the form:

$$\Phi_{DV}(d,l) = \int_{\Sigma(z>0)} ds\, (\mathbf{P}_S \cdot \mathbf{n})\varphi_{DE} - \frac{\varepsilon}{8\pi R_d^2}\int_{z>0} dv\, \varphi_{DE}^2 = \begin{cases} \dfrac{16\pi^2 P_S^2}{3\varepsilon}\dfrac{d^4}{l}(\ln(2l/d)-1), & R_d \to \infty \\ \dfrac{4\pi^2 P_S^2}{\varepsilon}d^2 R_d, & R_d \to 0 \end{cases}$$

The [1/1] Pade approximation for the "bulk" depolarization field energy $\Phi_{DV}(d,l)$ over variable $R_d$, acquires the following form:

$$\Phi_{DV}(d,l) \approx \frac{16\pi^2 P_S^2}{3\varepsilon}\frac{d^4 R_d (\ln(2l/d)-1)}{l R_d + 4d^2(\ln(2l/d)-1)/3} \tag{C.5}$$

Similarly to (C.5), it is easy to obtain from the Gauss theorem, that the expression for the "butt" depolarization energy $\Phi_{DS}(d,l)$ should be rewritten as:

$$\begin{aligned}\Phi_{DS}(d,l) &= \int_{\Sigma(z>0)} ds\,(\mathbf{P}_S \cdot \mathbf{n})\varphi_{DS} - \int_{\substack{(x^2+y^2)\le d^2 \\ z=0}} dxdy\, \varphi_{DS} P_S - \frac{\varepsilon}{8\pi R_d^2}\int_{z>0} dv\, (\varphi_{DS}^2 + 2\varphi_{DS}\varphi_{DE}) = \\ &= \int_{V_\Sigma} dv P_S \frac{d\varphi_{DS}}{dz} - \frac{\varepsilon}{8\pi R_d^2}\int_{z>0} dv\,(\varphi_{DS}^2 + 2\varphi_{DS}\varphi_{DE})\end{aligned} \tag{C.6}$$

Using (B.8), it is easy to estimate the following integrals in (C.6) with accuracy $O(\exp(-l/R_d)(d^2/l^2))$. It is follows from (C.6) and (C.7) that:

$$\Phi_{DS}(d) = \begin{cases} \dfrac{16\pi^2 P_S^2}{3\varepsilon} d^3 \left(\dfrac{4}{\pi} - (1+2\ln 2)\dfrac{d}{l}\right), & R_d \to \infty \\ \dfrac{4\pi^2 P_S^2}{\varepsilon} d^2 R_d, & R_d \to 0 \end{cases} \tag{C.7}$$

The [1/1] Pade approximation for the butt depolarization field energy $\Phi_{DS}(d,l)$ over variable $R_d$, acquires the following form:

$$\Phi_{DS}(d) = \frac{16\pi^2 P_S^2}{3\varepsilon} d^3 \frac{R_d}{\pi R_d/4 + 4d/3} \tag{C.8}$$